\documentstyle[pre,aps,multicol,epsfig]{revtex}

\begin{document}

\draft

\newcommand{\be}{\begin{eqnarray}}
\newcommand{\ee}{\end{eqnarray}}

\title{Systematic weakly nonlinear analysis of interfacial instabilities in Hele--Shaw flows}

\author{
E. Alvarez--Lacalle, J. Casademunt and J. Ort\'{\i}n. }
\address{
 Departament d'Estructura i Constituents de la Mat\`eria\\
Universitat de Barcelona,
Av. Diagonal, 647, E-08028 Barcelona, Spain}
\date{\today}
\maketitle

\begin{abstract}
We develop a systematic method to derive all orders of mode
couplings in a weakly nonlinear approach to the dynamics of the
interface between two immiscible viscous fluids in a Hele--Shaw
cell.  The method is completely general.  It includes both the
channel geometry driven by gravity and pressure, and the radial
geometry with arbitrary injection and centrifugal driving.  We
find the finite radius of convergence of the mode--coupling
expansion.  In the channel geometry we carry out the calculation
up to third--order couplings, which is necessary to account for
the time--dependent Saffman--Taylor finger solution and the case
of zero viscosity contrast.  Both in the channel and the radial
geometries, the explicit results provide relevant analytical
information about the role that the viscosity contrast and the
surface tension play in the dynamics.  We finally check the
quantitative validity of different orders of approximation against
a physically relevant, exact time--dependent solution.  The
agreement between the low order approximations and the exact
solution is excellent within the radius of convergence, and
reasonably good even beyond that.
\end{abstract}


\section{Introduction}

The morphological instability of fluid interfaces in Hele--Shaw flows
\cite{Bensimon86,Saffman86} has become a paradigm of interfacial
pattern formation in nonequilibrium systems
\cite{Pelce88,LangerlesHouches,Kesslerkop88}.  As opposed to the most
commonly studied ``bulk'' pattern forming systems \cite{Crosshoh93},
the inherent difficulties of the free--boundary problems associated to
interfacial growth makes the latter even more elusive to analytical
treatment.  As a prototype of interfacial instabilities in
diffusion--limited growth problems (including for instance dendritic
growth, solidification of mixtures, chemical electrodeposition, flame
propagation, etc.) the Saffman--Taylor problem \cite{Saffman58} is a
relatively simple case, both theoretically and experimentally,
well suited to gain insight into generic dynamical features in the
broad context of nonlinear interface phenomena.

The intrinsic difficulty of free--boundary problems is manifest in
the fact that the interface dynamics is highly non--local.
Furthermore, the nature of the instability (except in some cases,
such as in directional solidification of binary alloys) usually
produces non--saturated growth, which inevitably results in a
highly nonlinear dynamics.  In bulk instabilities, when the
control parameter is near threshold, the traditional weakly
nonlinear techniques lead to a universal description of patterns
in terms of amplitude equations, based on center manifold
reduction \cite{Crosshoh93,Manneville90}.  These techniques,
however, are not so useful for interfacial problems in which
nonlinearities do not saturate the growth.  This is the case for
instance of viscous fingering.  In the case of the channel
geometry (Saffman--Taylor problem) \cite{Bensimon86,Saffman58} the
interface restabilizes in a nontrivial morphology (the
Saffman--Taylor finger) which keeps growing at a finite rate.  For
circular geometries \cite{Paterson81,JDChen89} the patterns do not
reach an equivalent steady state and the interplay of
tip--splitting events and screening effects may result in a
variety of complicated morphologies.  In these problems all the
weakly nonlinear techniques apply only to a transient in the early
nonlinear regime.  Nevertheless, compared to the more traditional
ones \cite{Crosshoh93,Manneville90}, the weakly nonlinear analysis
developed in this paper is not restricted to situations near the
instability threshold, where a separation of scales is exploited.
Instead, we expand on the amplitudes of the whole spectrum of
modes.

In this paper we will deal both with channel and radial geometry
Hele--Shaw flows.  In the traditional Saffman--Taylor problem
(channel geometry) the pressure-- and gravity--driven
instabilities can be formally mapped into each other in the
appropriate reference frames, so there is really no different
interface dynamics for the two physical situations.  The problem,
then, contains two independent dimensionless parameters, namely, a
dimensionless surface tension $B$ and the viscosity contrast or
Atwood ratio $A$ \cite{Trygvason83}.  In the radial geometry,
though, there is no such formal mapping.  We will see that the
injection and the centrifugal forcing are not equivalent and three
independent parameters must be considered.

The situation most commonly studied in the literature is the high
viscosity contrast limit, $A=1$, where one of the two fluids is
non--viscous \cite{Bensimon86} (typically air displacing a viscous
fluid).  The singular perturbation character of the surface
tension $B$ has received most of the attention as responsible for
the subtle mechanism of {\it steady--state selection}, namely the
fact that surface tension ``selects'' a single finger solution out
of a continuum of solutions for $B=0$
\cite{Hong86,Shraiman86,Combescot86}.  More recently, the crucial
role of surface tension in the \it dynamics \rm of fingering
patterns has been pointed out.  Siegel and Tanveer
\cite{Tanveer93,Siegelprl96,Siegeljfm96} have shown that the
exact, nonsingular time--dependent solutions known for the case
with $B=0$ may differ significantly from the corresponding ones
with $B \rightarrow 0^{+}$ after a time which is of order one
($B^{0}$).  In practice, this implies that exact solutions of the
problem with $B=0$ (including those with no finite time
singularities) may lead to completely incorrect asymptotic
behaviour as compared to the regularized ones, with $B\ll 1$.  A
careful analysis of these questions may be found in
Ref.\cite{Magdaleno00}.  Notice however that such analysis
restricts to small $B$, while in many cases (for instance, for
fingers emerging naturally from the linear instability, with the
characteristic length scale of the linearly most unstable mode)
the effective dimensionless surface tension is necessarily $B \sim
1$.  Understanding the dynamics of finger competition in typical
experimental conditions thus requires considering relatively large
values of $B$, for which the perturbative techniques of
\cite{Tanveer93,Siegelprl96,Siegeljfm96} fail.

On the other hand, an important role of viscosity contrast $A$
in the dynamics of finger competition has been observed both numerically
\cite{Trygvason83,Trygvason85} and experimentally
\cite{Maher85,Difranjima89,Difranjimb89}.  A careful characterization
of the interface evolution has shown that for $A=0$ the finger
competition process is ineffective, and that the system does not
approach the usual single finger Saffman--Taylor attractor
\cite{Casademunt91,Casademunt94}.  Although the nature or existence
of other attractors is still an open question, it seems that the
basin of attraction of the Saffman--Taylor solution does depend on
$A$, and is particularly sensitive to $A$ in the neighborhood of $A
\simeq 1$.  In any case, it is clear that the viscosity contrast
plays also a crucial role in the highly nonlinear regime, and that
tuning $A$ in its full range is necessary to elucidate some of the
important open questions.

Finally, an interesting interplay between $B$ and $A$ in
connection with the selection problem is apparent in that, despite
the fact that single finger stationary solutions of any width do
exist for $B=0$ regardless of viscosity contrast $A$, the only
single finger time dependent solution of the $A=1$ ($B=0$) problem
which is also a solution for any viscosity contrast $A$ is the one
that fills one half of the channel \cite{Francessos1,Folch00},
which is precisely the solution selected by surface tension in the
limit $B \rightarrow 0$.  Whether deeper consequences concerning
the selection problem can be drawn from this fact is also an
interesting open question.

In order to gain analytical insight into these dynamical questions we
propose here a systematic weakly nonlinear expansion of the problem of
viscous fingering in Hele--Shaw flows, including all traditional setups
and the most recent one of rotating flows \cite{Carrillo96}.  The basic
motivation is to be able to extract information which is nonperturbative
in any of the two basic parameters, which are taken as completely
arbitrary.  The expansion parameter will be basically the mode
amplitudes.

In this paper we will focus on unstably stratified flows, for which
the approach is necessarily restricted to the early evolution of the
interface.  Although some of the nontrivial dynamic effects mentioned
above are associated to the highly nonlinear regime, it may be useful to
know within a controlled approximation to what extent those or other
effects show up already at the early stages of nonlinear mode coupling.

For the stably stratified case the weakly nonlinear analysis is
obviously valid for long times since all mode amplitudes decay with
time.  Although this configuration may seem trivial, this is not the
case in some situations, for instance when some external source
systematically drives the interface out of its equilibrium state (planar
or circular).  An example of this is the presence of noise sources, such
as the quenched noise associated to a porous medium \cite{porous}.  In a
study of the long--wavelength, low--frequency scaling properties of the
interface fluctuations, the knowledge of the lowest order nonlinear
terms and their dependence on parameters such as viscosity contrast is
crucial.  In this context the weakly nonlinear expansion is the starting
point of any Renormalization Group analysis of the relevant terms and of
the fixed points of the problem.  This line of research is clearly
beyond the scope of this paper and will not be pursued here.

The basic ideas of weakly nonlinear analysis developed here were
first applied to viscous fingering by Miranda and Widom
\cite{Mirandac98,Mirandar98} to both channel and circular geometry
(with fluid injection).  The present work is in part an extension
of those previous contributions in several directions, and in part
a detailed study of selected particular situations to assess the
validity and limitations of the approach.

First of all, we provide a fully systematic methodology which may be
carried out to arbitrary order.  We explicitly calculate the results
up to third order to include important situations for which the
second order couplings, first discussed in
Refs.\cite{Mirandac98,Mirandar98}, vanish identically, such as for
$A=0$ or for configurations with up--down symmetry (which includes the
physically relevant time dependent single finger solution with width
one half).  We give the explicit predictions both in real and Fourier
space.

In our formulation we also address the case of centrifugal forcing
of Hele--Shaw flows \cite{Schwartz89}.  The experimental study of
rotating Hele--Shaw flows has revealed a rich variety of new
phenomena \cite{Carrillo96,Carrillo99,Carrillo00}.  From a
theoretical point of view, new classes of exact solutions with
$B=0$ have been found \cite{Entov96,Crowdy1}.  The role of
rotation in the possible suppression of finite time cusp
singularities in the absence of surface tension has been discussed
in \cite{Rocco00}.  From an experimental point of view, important
differences in pattern morphology and new dynamical effects have
been found for low viscosity contrasts \cite{Alvarez01}.  It seems
thus important to have this case included in the weakly nonlinear
formalism.

In addition, our study extends the earlier results of Miranda
and Widom \cite{Mirandac98,Mirandar98} with the discussion of the
convergence of the weakly nonlinear analysis.  We find the explicit
exact criterion to assure uniform convergence of the series.  Beyond
that condition the series is asymptotic and different resummation
schemes are also explored.  The different orders of approximation,
including possible resummations, are carefully compared to exact
solutions for the case of a single finger configuration.  We find that
in some cases the agreement even at relatively low orders is quite
remarkable.

The layout of the rest of the paper is as follows:  in Section
\ref{Sec:2} we introduce the formalism.  Section \ref{Sec:3} deals with
the derivation of the weakly nonlinear equations and their application
to Hele--Shaw flows in channel geometry.  The analysis of the radial
geometry is carried out in Section \ref{Sec:4}, and Section \ref{Sec:5}
presents a numerical analysis of exact and approximate solutions.  The
main results and the conclusions are summarized in Section
\ref{conclusions}.

\section{Vortex sheet formalism}
\label{Sec:2}

\subsection{Channel Geometry}

\begin{figure}
\begin{center}
\epsfig{file=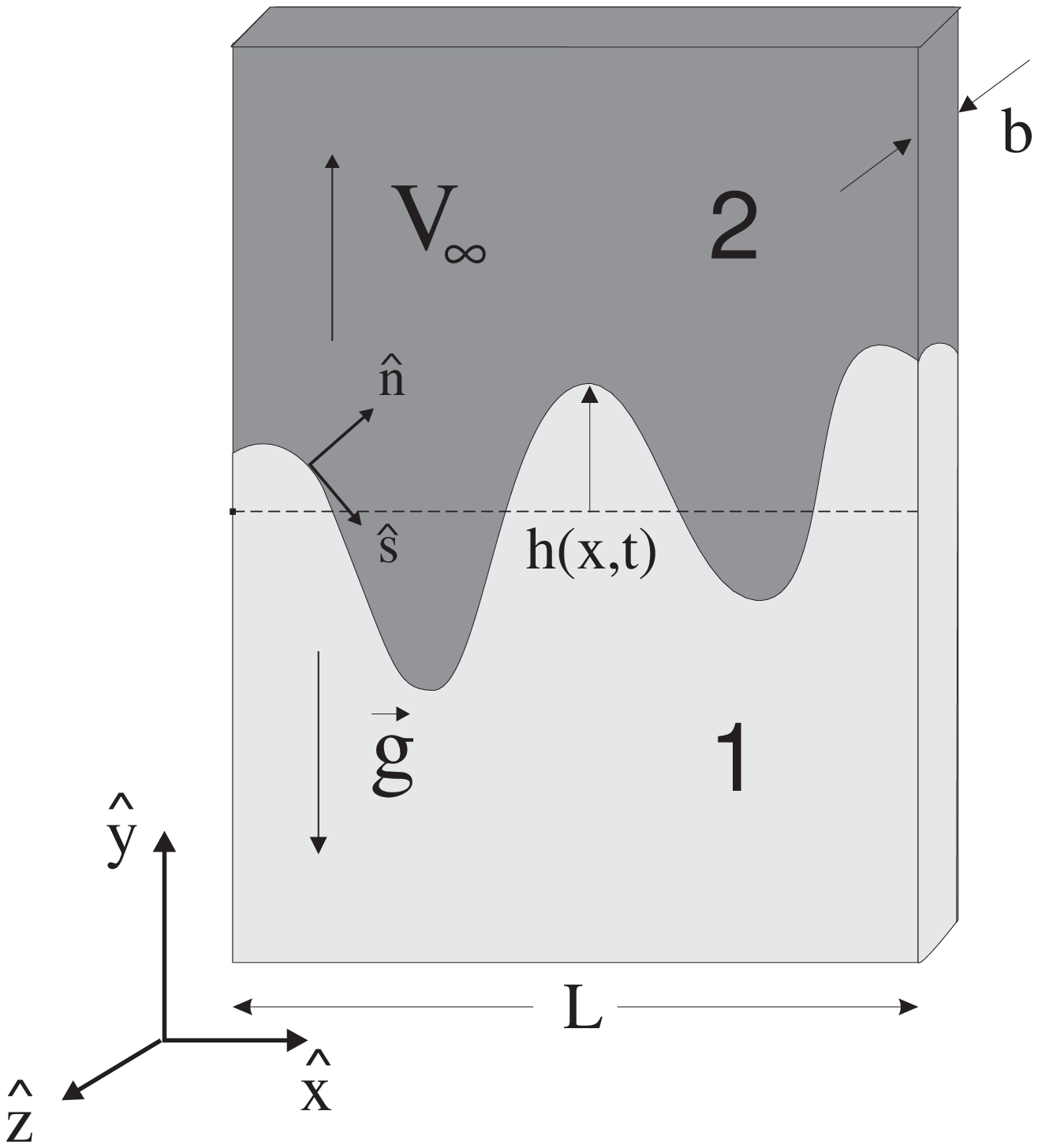,width=6cm} \vspace{0.1cm} \caption{Sketch
of the Hele-Shaw cell in channel geometry.} \label{Figcanal}
\end{center}
\end{figure}

Let us first consider the Hele--Shaw problem in the channel
geometry.  We consider fluid 1 (viscosity $\mu_{1}$, density
$\rho_{1}$) below fluid 2 ($\mu_{2},\rho_{2}$) (Fig.\
\ref{Figcanal}).  The $\hat{z}$ axis is perpendicular to the cell.
A velocity $V_{\infty}$ is imposed at infinity in the $\hat{y}$
direction.  Gravity points from fluid 2 to fluid 1.  The width of
the cell is $L$, the gap between plates is $b$, and the surface
tension between the fluids is $\sigma$.

The equations of motion for the interface and the boundary
conditions are well known \cite{Saffman58}.  Here we will use the
formulation of Tryggvason and Aref \cite{Trygvason83}.  We
introduce the velocity $\vec{U}=(\vec{u}_{1} + \vec{u}_{2})/2 $ as
the mean of the two limiting values of the velocities from both
sides of the interface ($\vec{u}_{1}, \vec{u}_{2}$) at a given
point.  This velocity $\vec{U}$ can be expressed in terms of the
vortex sheet distribution $\gamma$ at the interface as:
\be
\vec{U}(s,t)=\frac{1}{2\pi} P \int  \frac{\hat{z} \times
\left( \vec{r}(s,t)-\vec{r}(s^{\prime},t) \right) }{\mid
\vec{r}(s,t) -  \vec{r}(s^{\prime},t) \mid ^{2}} \gamma (s^{\prime} ,
t) ds^{\prime},
\label{Tr1}
\ee
where:
\be
\gamma = 2A({\bf \hat{U} \cdot \hat{s}}) + 2C{\bf \hat{y} \cdot
\hat{s}} + 2D \kappa _{s}, \label{2}
\ee
with $s$ the arclength, $\kappa$ the curvature, and
\be
A = \frac{\mu_{2} - \mu_{1}}{\mu_{2} + \mu_{1}},  \ \ \ C=\frac{g
b^{2} (\rho_{2} - \rho_{1})}{12(\mu_{2} + \mu_{1})} + AV_{\infty},
\ \ \ D=\frac{ \sigma b^{2}}{12(\mu_{2} + \mu_{1})}, \ \ \ \gamma
= (\vec{u}_{1} - \vec{u}_{2})\cdot\hat{s}.  \label{cons} \ee For
the purposes of this work it is convenient to rewrite these
equations in terms of the interface height $h(x)$, following
\cite{Goldstein98}.  The equations read:
\be
\vec{U}(x,t) = \frac{1}{2\pi} P \int_{-\infty}^{+\infty} \frac{(
h(x^{\prime}) -  h(x) , x - x^{\prime}  ) }{ (x - x^{\prime})^{2} +
( h(x^{\prime})-h(x) )^{2} } \tilde{\gamma}(x^{\prime}) dx^{\prime},
\label{tot1}
\ee
\be
\tilde{\gamma} = 2D\kappa_{x} + 2C h_{x} + 2A \vec{U} \cdot ( 1 ,
h_{x} ), \label{gm} \ee where:
\be
\tilde{\gamma}= \gamma \sqrt{ 1 + h_{x}^2 }  , \qquad \kappa = \frac{
h_{xx}}{( 1 +  h_{x}^{2} ) ^{3/2} }.
\label{ka}
\ee
The dependence of $h$ and $\tilde{\gamma}$ on time is not written
explicitly.

To complete the definition of the moving boundary problem the
continuity of the normal velocity at the interface is required.
This means that the velocity in the $y$ direction, $dh/dt$,
projected along the normal direction, is equal to the normal
component of the average velocity of the interface, $\vec{U} \cdot
\hat{n}$:
\be
\frac{dh}{dt} = U_{\hat{y}} -  U_{\hat{x}} h_{x}.
\label{tot3}
\ee
In Section \ref{Sec:3} we will consider  the interface in the
comoving frame ($\overline{h}=0$) and will look for the equation of
evolution of $h(x,t)$.

\subsection{Radial geometry}

\begin{figure}

\begin{center}
\epsfig{file=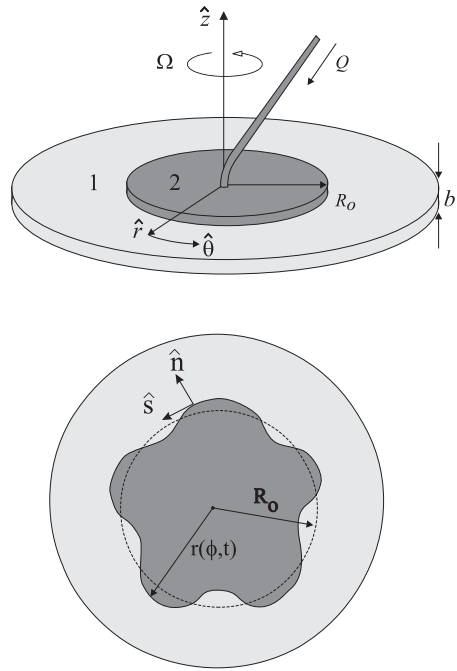,width=6cm} \vspace{0.5cm} \caption{Sketch
of the Hele-Shaw cell in circular geometry.} \label{Figradial}
\end{center}
\end{figure}

We proceed to write the corresponding equations for a circular
cell rotating with angular velocity $\Omega$.  The initial
condition is $ R=R_{0}$ (constant) and we label the outer (inner)
fluid as the fluid 1 (2) which have known viscosities $\mu_1$,
$\mu_2$ and densities $\rho_1$, $\rho_2$ (Fig.\ \ref{Figradial}).

If we perform a change to polar coordinates in (\ref{Tr1}), and
from arclength $s$ to angle $\phi$, we get for the mean velocity:
\be
U_{\hat{r}}=\vec{U}(\phi_{1},t) \cdot \hat{r}=\frac{1}{2\pi} P
\int_{0}^{2\pi} \frac{r_2^2 \sin(\phi_2- \phi_1)}{r_1^2
+r_2^2-2r_1r_2 \cos(\phi_2-\phi_1)} \tilde{\gamma} (\phi_2) d\phi_2,
\label{imp}
\ee
\be
U_{\hat{\phi}}=\vec{U}(\phi_{1},t) \cdot \hat{\phi}=\frac{1}{2\pi} P
\int_{0}^{2\pi} \frac{r_1r_2 - r_{2}^{2}\cos(\phi_2- \phi_1)}{r_1^2
+r_2^2-2r_1r_2 \cos(\phi_2-\phi_1)}\tilde{\gamma} (\phi_2)  d\phi_2,
\label{impd}
\ee
where $\tilde{\gamma} =
\sqrt{1+(r_{\phi}/r)^2} (\vec{u}_1-\vec{u}_2)\cdot\hat{s}$, and we have
used the notation $r(\phi_{1},t) \equiv r_1$,
$r(\phi_{2},t) \equiv r_2$.

In the presence of sinks or sources the velocities $\vec{u}_{1}$ and
$\vec{u}_{2}$ which define $\vec{U}$ include only the solenoidal
part of the total velocity field.  For this reason, in the
presence of injection ($Q>0$) or withdrawal ($Q<0$) of the inner
fluid, Eqs. (\ref{imp}) and (\ref{impd}) must be supplemented with
the corresponding irrotational part of the velocity field.

In order to obtain the expression for the vorticity as a function of
$\vec{u}_{i}$, we use the local equations known for this problem
\cite{Carrillo96,Schwartz89}:
\be
\vec{\nabla}p_i=-\frac{12\mu_i}{b^2} \left(\vec{u}_i + \frac{Q}{2\pi
r}\hat{r}\right) + \Omega ^2 \rho_i r \hat{r}, \qquad i=1,2
\ee
and the boundary conditions:
\be
p_2 - p_1 = \sigma \kappa, \qquad \qquad u_{\hat{n}}^{(1)} =
u_{\hat{n}}^{(2)}.
\ee
After some algebra, we can write an expression for the vorticity as a
function of $\vec{U}$:
\be
\tilde{\gamma}=\frac{b^2}{12}\frac{2\sigma}{r(\mu_1 +
\mu_2)}\kappa_{\phi} + 2A \vec{U}\cdot (\frac{r_{\phi}}{r} , 1 ) + 2A
\frac{Q}{2\pi r^2}r_{\phi} - \frac{b^2}{12} \frac{ 2 \Omega ^2(\rho_2
- \rho_1)}{\mu_1+\mu_2}r_{\phi}, \label{gmvip} \ee with
\be
\kappa
=\frac{(r^2+2r_{\phi}^2-rr_{\phi\phi})}{(r^2+r_{\phi}^2)^{\frac{3}{2}}},
\qquad  A=\frac{\mu_2-\mu_1}{\mu_1+\mu_2}.
\ee

The last step is to derive the equation of the continuity of the
normal velocity.  Now the projection of the radial velocity along
the normal direction $\hat{n}$ has two contributions: the
solenoidal part of the average velocity, $\vec{U}$,
projected along $\hat{n}$, and the irrotational part of the mean interface
velocity, $\vec{U}_{irrot}$:
\be
\frac{dr}{dt} \hat{r} \cdot \hat{n} = \vec{U} \cdot \hat{n} +
\vec{U}_{irrot} \cdot \hat{n}  \ \ \rightarrow \ \
\frac{dr}{dt}=U_{\hat{r}} - \frac{r_{\phi}}{r}U_{\hat{\phi}} +
\frac{Q}{2 \pi r}. \label{nvir} \ee

This completes the system of equations for the two
geometries considered.  For an interface with no overhangs, Eqs.
(\ref{tot1}), (\ref{gm}), and (\ref{tot3}) are the starting point
for the generalized weakly nonlinear analysis in the rectangular
cell. Equations (\ref{imp}), (\ref{impd}), (\ref{gmvip}), and
(\ref{nvir}) play the same role for the analysis in
the rotating circular cell.

\section{Systematic weakly nonlinear analysis.  Channel geometry}
\label{Sec:3}

Our goal in this section is to introduce a systematic method to
derive an evolution equation of the interface in real space to a
given order in nonlinear couplings, in the channel geometry.  The
different orders of mode couplings will be ordered as powers of a
``book--keeping'' parameter $\varepsilon$, to be defined below.  The
evolution of the interface will thus take the form:
\be
\frac{dh}{dt} = F[h] + \varepsilon G[h] + \varepsilon ^{2} I[h] +
\ldots, \label{generity} \ee where $F[h], G[h], \ldots$ are
nonlocal operators on the function $h(x,t)$, including
nonlinearities of order $n+1$ in the term of order $\varepsilon
^{n}$.  The small parameter $\varepsilon$ is defined as the ratio
of two lengths, $\varepsilon=w/L$.  We take $w$ as a measure of
the characteristic scale of variation of the interface $h(x)$,
while $L$ is either the width of the cell or, alternatively, the
characteristic scale of variation in the $x$ direction.  The
weakly nonlinear regime is defined by the condition $w \ll L$.

The order $\varepsilon^0$ in Eq.  (\ref{generity}) corresponds to
the linearized equation.  The order $\varepsilon^1$, when written
in Fourier space, corresponds to the result of Miranda and Widom
\cite{Mirandac98}.  Here we will perform the explicit calculation
to one order higher in $\varepsilon$ (up to $I[h]$) which is the
leading nonlinear contribution in several important cases, such as
the one discussed in Section \ref{subsecdit}.  The systematics is
presented in the channel geometry for simplicity.

\subsection{Dimensionless equations, expansion and convergence}
\label{SectionIIIA}

We scale the interface height with $w$, the coordinate $x$ with $L$,
and the time with $L/C$, where the velocity $C$ has been defined in
(\ref{cons}).

The equations (\ref{tot1}), (\ref{gm}), (\ref{ka}), and
(\ref{tot3}) become:
\be
\vec{U}(x,t) = \frac{1}{2\pi} P \int_{-\infty}^{+\infty}
\frac{(\varepsilon \left[h(x^{\prime}) -  h(x)\right] , x -
x^{\prime} ) }{ (x - x^{\prime})^{2} \left\{ 1 + \varepsilon^{2}
\left[\frac{  h(x^{\prime})-h(x)}{x^{\prime}-x}\right] ^2 \right\}}
\tilde{\gamma}(x^{\prime}) dx^{\prime}, \label{gete}
\ee
\be
\tilde{\gamma} = 2B\kappa_{x} + 2 \varepsilon h_{x} + 2A \vec{U}
\cdot ( 1 , \varepsilon h_{x} ), \label{gmvera}
\ee
where
\be
B =\frac{\sigma b^{2}}{12CL^{2}( \mu_1+\mu_2)}, \ \ \ \kappa =
\frac{\varepsilon h_{xx}}{( 1 + \varepsilon ^{2} h_{x}^{2} ) ^{3/2}
},
\ee
and
\be
\frac{dh}{dt} = \frac{1}{\varepsilon} U_{\hat{y}} - U_{\hat{x}}
h_{x}. \label{dhdt}
\ee

The starting point of our approach is an expansion of $\vec{U}$,
$\tilde{\gamma}$ and $\kappa$ in powers of $\varepsilon$.  Notice
that the term
$\left[ h(x^{\prime})-h(x) \right] ^2 \ \left[
x^{\prime}-x \right] ^2$ between curly
brackets in (\ref{gete}) is bounded provided that $h_x$ does not
diverge.  Equation (\ref{dhdt}) thus takes the form:
\be
\frac{dh}{dt} = \frac{1}{\varepsilon}U_{\hat{y}}^{(0)} +
U_{\hat{y}}^{(1)} -  h_{x} U_{\hat{x}}^{(0)} + \varepsilon
(U_{\hat{y}}^{(2)} -  h_{x} U_{\hat{x}}^{(1)}) + \cdots,
\label{devel}
\ee

Before pursuing the calculation in detail, let us point out that
the $\varepsilon$--expansion in Eq.(\ref{devel}) has a finite
radius of convergence.  This is guaranteed by the properties of
uniform convergence of both the expansion of the denominator in
Eq.(\ref{gete}) and the curvature.  These properties allow us
to commute the expansion with the integral in Eq.(\ref{gete}) and
yield a convergent series of the form (\ref{devel}).  The
radius of convergence of the expansion of the denominator of
Eq.(\ref{gete}) is given by the condition
$\varepsilon^2
\left[ h(x^{\prime})-h(x) \right] ^2 \ \left[
x^{\prime}-x \right] ^2 < 1$,
while the convergence of the curvature expansion is assured by the
condition $\varepsilon^2 h^2_x < 1$.  In the nonscaled original
variables, the two conditions for convergence coincide, and read
$MAX(|h_x|)<1$.  If $|h_x|<1$ in the whole domain of integration,
then the $\varepsilon$--expansion converges.  If the condition is
not fulfilled in some intervals, then Eq.(\ref{devel}) is an
asymptotic expansion.  Even in this case, the expansion contains
useful information about the original problem.

An interesting case is $A=0$ which makes the vorticity independent
of $\vec{U}$ in Eq.(\ref{gete}), so that Eq.(\ref{dhdt}) becomes a
closed equation for $h(x,t)$.  Then, in Eq.(\ref{devel}) it is
easy to show that $U_{\hat{y}}^{(n)}=0$ when $n$ is even and
$U_{\hat{x}}^{(n)}=0$ when $n$ is odd, a property that makes the
even power terms of the expansion in Eq.(\ref{devel}) to vanish.

\subsection{First and second order expansion ($\varepsilon ^0$ and
$\varepsilon ^1$)}

Following the scheme introduced in the previous section we obtain
from Eq.(\ref{gete}) that:
\be
U_{\hat{x}}^{(0)} = 0, \qquad U_{\hat{x}}^{(1)} = \frac{1}{2\pi} P
\int_{-\infty}^{+\infty} \frac{  h(x^{\prime}) -  h(x)  }{ (x -
x^{\prime})^{2}} \tilde{\gamma}^{(0)}(x^{\prime}) dx^{\prime},
\ee
\be
U_{\hat{y}}^{(0)} = \frac{1}{2\pi} P \int_{-\infty}^{+\infty}
\frac{\tilde{\gamma}^{(0)}(x^{\prime})}{x - x^{\prime}}
dx^{\prime}, \qquad U_{\hat{y}}^{(1)} =  \frac{1}{2\pi} P
\int_{-\infty}^{+\infty} \frac{\tilde{\gamma}^{(1)}(x^{\prime})}{x
- x^{\prime}} dx^{\prime}.
\ee
Since $\tilde{\gamma}^{(0)}=2B\kappa_{x}^{(0)} + 2AU_{\hat{x}}^{(0)}=0$
we get $U_{\hat{x}}^{(1)}=U_{\hat{y}}^{(0)}=0$.  With this result we can
study the first order term in the vorticity equation:
\be
\tilde{\gamma} ^{(1)}(x) =2B\kappa_{x}^{(1)} + 2h_{x} +
2AU_{\hat{x}}^{(1)} + 2A h_{x}U_{\hat{y}}^{(0)}.
\ee
Taking into account the definition of the Hilbert Transform:
\be
H[f(x^{\prime})]= \frac{1}{\pi} P \int_{-\infty}^{+\infty}
\frac{f(x^{\prime})}{x^{\prime} - x} dx^{\prime}, \ee
Eq.(\ref{devel}) up to order $\varepsilon ^0$ reads:
\be
\frac{dh}{dt} =
- H[Bh_{x^{\prime}x^{\prime}x^{\prime}} +
h_{x^{\prime}}].  \label{deht}
\ee
The linear operator $F[h]$ in Eq.(\ref{generity}) thus reads explicitly:
\be
F[h] = \frac{1}{\pi} P \int_{-\infty}^{+\infty}
\frac{\left(h + Bh_{x^{\prime}x^{\prime}}\right)_{x^{\prime}} }{x -
x^{\prime}} dx^{\prime} .
\ee

Writing $h(x,t)$ as a superposition of Fourier
modes in Eq.(\ref{deht}) we recover the linear dispersion relation:
\be
\frac{\dot{\delta} _{k}(t)}{\delta _{k}(t)} = \lambda (k) = \mid k
\mid (1 - Bk^{2}).
\ee
We will take $k$ as an integer but we should
keep in mind that, upon restoring dimensions, $k$ should
become $(2 \pi/L)n$, with $n$ integer.

Let us pursue the systematics of the method by computing the
next order in $\varepsilon$ in Eq.(\ref{dhdt}):
\be
\frac{dh}{dt} = U_{\hat{y}}^{(1)} + \varepsilon  U_{\hat{y}}^{(2)}
= F[h] + \varepsilon G[h].
\ee
The computation of $U_{\hat{y}}^{(2)}$ requires an expression of
the vorticity up to second order.  This includes the
evaluation of $U_{\hat{x}}^{(2)}$, which must be computed from the
first order term of the vorticity.  We get:
\be
U_{\hat{x}}^{(2)}= H\left[ \left(h(x^{\prime})
f(x^{\prime})\right)_{x^{\prime}}\right] -
h(x)H[f_{x^{\prime}}(x^{\prime})], \label{unavez}
\ee
with $f(x) \equiv (Bh_{xx} +h)_{x}=\tilde{\gamma}^{(1)}(x)/2$, and:
\be
U_{\hat{y}}^{(2)} = A \left\{ H \left[h_{x^{\prime}}
H[f(x^{\prime\prime})]\right] + H \left[ h(x^{\prime})
H[f_{x^{\prime\prime}}(x^{\prime\prime})]\right] +
(hf)_{x}\right\}.
\ee
Taking the Fourier transform, we obtain:
\be
\dot{\delta}_k  = \lambda (k) \delta_k  + \varepsilon A \mid k
\mid \sum_{s=-\infty}^{+\infty}\left[1 - \mbox{sgn}(ks) \right]
\lambda (s) \delta_s (t) \delta_{k-s} (t),
\label{secmod}
\ee
which coincides with the result of Miranda and Widom in Ref.
\cite{Mirandac98}.

\subsection{Third order expansion ($\varepsilon ^2$)}

The expansion to order $\varepsilon^2$ is necessary to account for
the lowest order nonlinearities in the case of zero viscosity
contrast, and for other relevant situations such as the time
dependent Saffman--Taylor finger solutions (section
\ref{subsecdit}).  We now have:
\be
\frac{dh}{dt} = {\cal{O}}(\varepsilon^0) +
{\cal{O}}(\varepsilon^1) + \varepsilon ^2 (U_{\hat{y}}^{(3)} -
h_xU_{\hat{x}}^{(2)} ) + \cdots \ee We have already computed
$U_{\hat{x}}^{(2)}$ in Eq. (\ref{unavez}).  On the other hand:
\be
U_{\hat{y}}^{(3)}= - \frac{1}{2} H[
\tilde{\gamma}^{(3)}(x^{\prime})] + \frac{1}{2 \pi} P
\int_{-\infty}^{+\infty} \frac{ \left[h(x^{\prime}) -
h(x)\right]^2 }{ ( x^{\prime}-x)^{3}} \tilde{\gamma}
^{(1)}(x^{\prime}) dx^{\prime}. \label{uyt} \ee Integrating twice
by parts, the last integral can also be written as a Hilbert
Transform.  After some algebra we obtain the explicit form of the
operator $I[h]$ containing the cubic nonlinearities in
Eq.(\ref{generity}), which reads:
\be
\begin{array}{ll}
I[h] = & \frac{3}{2}H[ g(x^{\prime})]  +
\frac{1}{2}H \left[ \left( f(x^{\prime}) [ h(x)-h(x^{\prime})]^2
\right) _{x^{\prime}x^{\prime}}\right] \\
& -h_x H\left[\left( f(x^{\prime}) [h(x^{\prime}) - h(x)]
\right) _{x^{\prime}}\right] + V[h,A]
\end{array}
\ee
with:
\be
V[h,A]=A^2 H \left[h(x^{\prime})H
[\tau_{x^{\prime\prime}}(x^{\prime\prime})] +
h_{x^{\prime}}(x^{\prime})H[\tau(x^{\prime\prime})]\right] + A^2
(h\tau)_{x},
\ee
and:
\be
g(x)=B\left(h_{xx}h_x^2\right)_x, \qquad \tau(x)=U_{\hat{x}}^{(2)} +
h_xU_{\hat{y}}^{(1)}=\frac {\tilde {\gamma}^{(2)}(x)} {2A}.
\ee

The same result in Fourier space takes the form:
\be
\dot{\delta} _{k}(t)= {\cal{O}}(1,\varepsilon) + \varepsilon^2
\sum_{s,l=-\infty}^{+\infty} \delta_{l} \delta_{s-l}\delta_{k-s}
\left[A^2 T(k,s,l) - \frac{3}{2}B Y(k,s,l) + W(k,s,l)\right],
\label{triorder}
\ee
with:
\be
T(k,s,l)=\mid k \mid \mid s \mid \lambda (l) \left[1 -
\mbox{sgn}(ks)\right] \left[1 - \mbox{sgn}(ls)\right],
\ee
\be
Y(k,s,l)=\mid k \mid l^2(s-l)(k-s),
\ee
\be
W(k,s,l)=\left[l\left(\frac{l}{2}+k-s\right) -s k \mbox{ sgn}(ls) +
\frac{k^2}{2}  \mbox{ sgn}(kl)\right] \lambda(l).
\ee

It is clear that the viscosity contrast in Eq.(\ref{triorder})
appears squared because of the reflection symmetry  (the
simultaneous change $A \rightarrow -A$ and $h \rightarrow -h$ is a
dynamical symmetry of the problem).  Symmetry reasons alone,
however, do not allow to discard a three mode coupling
contribution when $A=0$.  We see from our calculation that
a three mode coupling is indeed present independently of $A$.

Following this scheme, the fourth order will carry a contribution
proportional to $A$, and another proportional to $A^{3}$, for
symmetry reasons.  The fifth order will carry a contribution
independent of $A$ and two others, proportional to $A^2$ and
$A^4$, respectively.  This scheme will continue for subsequent
orders.

\subsection{Analysis of the time dependent single finger solution}
\label{subsecdit}

In this section we perform a detailed analysis of the weakly nonlinear
expansion in cases where exact solutions are known, namely single finger
configurations with $B=0$.  This allows to study how exact properties of
solutions show up at the different orders, particularly concerning the
role of viscosity contrast $A$.

At this point it is worth recalling that the case $A=1$ allows for
a continuum of Saffman--Taylor finger solutions corresponding to
different finger widths.  A continuum of time dependent exact
solutions, leading to those stationary states, is also known for
$A=1$.  However, for $A \neq 1$ only the one of width
$\lambda=\frac{1}{2}$ remains a solution, $\lambda$ being the
ratio of the finger width to the width of the channel.  This
result, which has been recently addressed in Ref.\cite{Folch00}
although it was first discovered in Ref.\cite{Francessos1}, points
out an intriguing connection between the width selection problem
and the dynamical role of viscosity contrast.  Here we will
analyze the interplay between $A$ and $\lambda$ in the early
nonlinear regime, and elucidate at what stage of the nonlinear
dynamics does the viscosity contrast $A \neq 1$ prevents the
possibility of having $\lambda\neq 1/2$.

From now on we consider $L=2\pi$, $B=0$, $C=1$, and we take as
scaling velocity $(1 - \lambda) C = (1 -
\lambda)\equiv\frac{\eta}{2}$.  Conformal mapping techniques
enable us to write the single finger solution of this problem in
the form \cite{Francessos1,Folch00}:
\be
f(w,t)=- \ln w + d(t) + \eta \ln (1 - \alpha(t) w), \label{mappi}
\ee where $f(w,t) = y + ix$, is an analytic function inside the
unit disk in the $w$--complex plane, which maps that disk into the
physical region occupied by the more viscous fluid.  The interface
is obtained in a parametric form by setting $w=e^{i \theta }$.
The functions $\alpha (t)$, $d(t)$ verify:
\be
\dot{d}(t)=\frac{\eta}{2-\eta}\left(\frac{\eta}{2} -
\frac{\dot{\alpha}(t)}{\alpha (t)}\right), \label{setun}
\ee
\be
\frac{\dot{\alpha}(t)}{\alpha (t)} = \frac{\eta} { 2 +
\eta(\eta-2) + \eta(2-\eta)\frac{1+\alpha^2(t)}{1-\alpha^2(t)} }.
\label{setdos}
\ee
To obtain an expression for $h(x)$ in the weakly nonlinear regime of
the evolution, $\alpha(t)$ and $d(t)$ are expanded in powers of a
small parameter $\nu$:
\be
d(t) = d^{(0)}(t) + \nu d^{(1)}(t) + \nu ^2 d^{(2)}(t)+ \cdots \ \
, \qquad \alpha (t) = \nu \alpha^{(0)} (t) + \nu ^2 \alpha ^{(1)}
(t).
\ee
Introducing these expansions in Eqs.(\ref{setun}) and
(\ref{setdos}) we obtain:
\be
\dot{d}^{(0)}=\dot{d}^{(1)}=0 , \qquad \dot{d}^{(2)}=
\frac{\eta^3}{2} (\alpha^{(0)})^2(t), \qquad \dot{d}^{(3)}= \eta
^3 \alpha^{(0)}\alpha^{(1)}, \ \ \cdots
\ee
\be
\dot{\alpha}^{(0)}=\frac{\eta}{2}\alpha^{(0)}, \qquad
\dot{\alpha}^{(1)}=\frac{\eta}{2}\alpha^{(1)}, \qquad
\dot{\alpha}^{(2)}=\frac{\eta}{2}\left(\alpha^{(2)} -\eta (2 -
\eta)(\alpha^{(0)})^3\right), \ \ \cdots \label{alfas} \ee which
will be useful later.  From (\ref{mappi}) and (\ref{setun}) we
obtain:
\be
\begin{array}{ll}
 y=h= &  -\nu \eta \alpha^{(0)} \cos \theta -
\nu^2 \left( \eta \alpha^{(1)} \cos \theta +
\frac{\eta}{2}(\alpha^{(0)})^2 \cos 2\theta - d^{(2)} \right) \\ & -
\nu ^3 \left( \eta \alpha^{(2)}\cos\theta + \eta
\alpha^{(0)}\alpha^{(1)}\cos 2\theta + \frac{\eta}{3}
(\alpha^{(0)})^3 \cos 3\theta -d^{(3)} \right)+ {\rm \cal{O}}(\nu^4)
, \label{hexp}
\end{array}
\ee
\be
x= -\theta -\nu \eta \alpha^{(0)} \sin \theta - \nu^2 \left( \eta
\alpha^{(1)} \sin \theta + \frac{\eta}{2}(\alpha^{(0)})^2 \sin 2
\theta \right) + {\rm \cal{O}}(\nu^3).
\ee
This last equation can be inverted in a systematic way to get:
 \be
\theta = -x + \nu \eta \alpha^{(0)} \sin x + \nu^2 \left( \eta
\alpha^{(1)} \sin x + \frac{\eta}{2} (1-\eta)(\alpha^{(0)})^2 \sin
2x \right) + {\rm \cal{O}}(\nu^3).
\ee
The relation between $\theta$ and $x$ can be inverted in Eq.(\ref{hexp})
and the cosine functions expanded.  We obtain in this way the following
expression for $h(x,t)$:
\be
\begin{array}{ll}
h(x,t)= & - \nu \eta \alpha^{(0)} \cos x - \nu^2 \eta\left( \alpha^{(1)}
\cos x + \frac{1-\eta}{2} (\alpha^{(0)})^2 \cos 2x\right) + \\
& \nu^3 \eta \left\{ (\eta-1) \alpha^{(0)} \alpha^{(1)} \cos 2x +
\left(\frac{3\eta}{8}(\eta-2)(\alpha^{(0)})^3 -
\alpha^{(2)}\right) \cos x + \right. \\
& \left. \left( \frac{3\eta}{8}(2-\eta) -
\frac{1}{3} \right) (\alpha^{(0)})^3 \cos 3x\right\} + {\rm
\cal{O}}(\nu^4) . \label{hexpan}
\end{array}
\ee

In order to follow the scheme developed in the previous section,
we must measure $h(x,t)$ in units of its characteristic amplitude
$\nu$.  In this way, the previous equation (\ref{hexpan}) takes
the form:
\be
h=h^{(0)} + \nu h^{(1)}+ \nu^2 h^{(2)} + {\rm \cal{O}}(\nu^3),
\label{simexp}
\ee
where $h(x,t)$ represents now $h(x,t)/ \nu$,
and the small parameter $\nu$ is directly comparable to the small
parameter $\varepsilon$ of the previous section.  The expression
for $h(x,t)$ can be regarded as a series of modes of decreasing
amplitude in our expansion (\ref{generity}) (written in the units
of this problem), and matching the corresponding powers in either
$\varepsilon$ or $\nu$ we obtain:
\be
\frac{dh^{(0)}}{dt}=\frac{\eta}{2}F[h^{(0)}]
\ee
to order $\varepsilon ^{0}$.

This identity can be verified (independently of $A$) by
substituting the expression of $h^{(0)}$ and using
Eqs.(\ref{alfas}) and (\ref{deht}) for the left and right hand
side respectively.  At order $\varepsilon$ we have:
\be
\frac{dh^{(1)}}{dt}=\frac{\eta}{2}\left(F[h^{(1)}] +
G[h^{(0)},h^{(0)}]\right) .
\ee
The second term of the right hand side gives no contribution, and we
obtain an interesting result:
the equation at order $\varepsilon$ is verified independently of
$\eta$ and $A$.  Hence, we cannot establish the difference between
$A=1$ (compatible with any $\eta$) and $A \neq 1$ (compatible only
with $\eta=1$) at this order.

The difference between these two situations arises at order
$\varepsilon^2$:
\be
\frac{dh^{(2)}}{dt}= \frac{\eta}{2} \left(F[h^{(2)}] +
G[h^{(0)},h^{(1)}] + L[h^{(0)},h^{(0)},h^{(0)}]\right).
\ee
Since the left hand side of the equation involves only the modes $\cos
x$, $\cos 2x$, and $\cos 3x$, the right hand side of the equation
must include these same modes with the same coefficients.  The
coefficients for $\cos 2x$ and $\cos 3x$ are easily matched.  For
$\cos x$ the left hand side reads:
\be
\left(\frac{\eta^3(\eta-2)}{16}(\alpha^{(0)})^3 -
\frac{\eta^2}{2}\alpha^{(2)}\right) \cos x,
\ee
where we have used Eq. (\ref{alfas}), and the right hand side reads:
\be
\left(\frac{3\eta^3(\eta-2)}{16}(\alpha^{(0)})^3
-\frac{\eta^2}{2}\alpha^{(2)}\right) \cos x +
A\frac{\eta^3(1-\eta)}{4}(\alpha^{(0)})^3 \cos x +
\frac{\eta^4}{8}(\alpha^{(0)})^3 \cos x.
\ee
Hence, the requirement for matching the coefficients on both sides is:
\be
\frac{\eta^3}{4}(1-\eta)=A \frac{\eta^3}{4}(1-\eta),
\ee
which is always true if $\eta=1$ ($\lambda=1/2$), but for $\eta \neq 1$
($\lambda \neq 1/2$) it requires $A=1$.  This result shows that
the nontrivial relationship between $A$ and $\lambda$ which is
known from exact solutions is already manifest at
the early nonlinear stages of the dynamics.  This clearly illustrates
the potential usefulness of the weakly nonlinear expansion at a purely
analytical level, in that a dynamical property
of the problem which must be satisfied at all orders (in this case the
incompatibility of $A \neq 1$ and $\lambda \neq 1/2$) may be
detected at a low order in the expansion, without having to know the
exact solution to all orders.

If we pursue the expansion to higher orders,
the general expression for $h^{(n)}$ takes the form:
\be
h^{(n)}=\sum_{k=1}^{n+1} \beta_{k}^{(n)} (t) \cos(kx),
\label{eshden} \ee where each coefficient is a function of $\alpha
(t)$.  The even modes of the solution with $\eta=1$ have zero
coefficients because they must remain invariant under a change of
sign and translation by $\pi$.

So far in this section we have restricted ourselves to
$B=0$ in order to compare with exact results.  However, we can carry out
the analysis also for $B \neq 0$.  It can be shown that the structure of
the expression (\ref{eshden}) is preserved in this case.  The
coefficients $\beta_{k}^{(n)} (t)$ are obtained as solutions of a set
of differential equations which contain the surface tension $B$.  For
instance, to third order we have:
\be
h(x,t) = \left(\beta_1^{(0)} + \varepsilon \beta_1^{(1)} +
\varepsilon^2 \beta_1^{(2)}\right) \cos x + \left( \varepsilon
\beta_2^{(1)} + \varepsilon^2 \beta_2^{(2)}\right) \cos 2x +
\varepsilon^2 \beta_3^{(2)} \cos 3x,
\label{generalo}
\ee
which, introduced in Eq.(\ref{generity}), leads to the following
equations:
\be
\begin{array}{c}
 \dot{\beta}_1^{(0)}= V_0(1-B)\beta_1^{(0)}, \qquad
\dot{\beta}_1^{(1)}= V_0(1-B)\beta_1^{(1)}, \qquad
\dot{\beta}_2^{(1)}= 2V_0(1-4B)\beta_2^{(1)},  \\ \displaystyle{
\dot{\beta}_2^{(2)}= 2V_0(1-4B)\beta_2^{(2)}, \qquad
\dot{\beta}_3^{(2)}= 3V_0(1-9B) \beta_3^{(2)} -
\frac{9BV_0}{8}(\beta_1^{(0)})^3, } \\
\displaystyle{\dot{\beta}_1^{(2)}= V_0(1-B)\beta_1^{(2)} + A
V_0(1-B)\beta_2^{(1)}\beta_1^{(0)} - \frac{V_0}{8}(2-5B)
(\beta_1^{(0)})^3,}  \label{eqbet}
\end{array}
\ee once the dimensions are reintroduced.  In this way we can also
see how surface tension perturbs the dynamics at the weakly
nonlinear regime.  Clearly, at these early stages of the nonlinear
evolution there is no sign of the singular perturbation character
of surface tension, which, at the late stages of the evolution
will be responsible for the selection of the steady
state\cite{Bensimon86}.

\section{Application of the method to the rotating Hele--Shaw cell}
\label{Sec:4}

\subsection{Dimensionless equations in the radial geometry}
\label{dimeq-in-rad}

The $\varepsilon$ expansion in the channel geometry originated in
the different scaling of lengths in the $x$ and $y$ directions.
To perform an equivalent expansion in the radial geometry we first
split the function $r(\phi,t)$ in two parts, namely the radius of
the unperturbed circle and the departure from this circle:
\be
r(\phi,t) \qquad \rightarrow \qquad \sqrt{R_0^2 + \frac{Qt}{\pi}}
+ r(\phi,t) = R(t) + r(\phi,t).
\ee
The largest length scale in the problem is given by $R(t)$, which is
only a constant in the
absence of injection.  For finite injection rate, $R(t)$ defines an
evolving (unstable) solution.  The proper scaling of
$r(\phi, t)$, which will define the weakly
nonlinear dynamics now, is not the naive extension of the scaling
defined in the channel geometry (namely $\varepsilon = w/R$, $w$ being
the typical departure from circularity).  In fact, this scaling would
mix different orders of the weakly nonlinear analysis.  The reason
for this was already pointed out by Miranda and
Widom \cite{Mirandar98}: in the radial geometry, mass
conservation implies that the zero mode (which in the channel geometry
is decoupled from the rest and drops out of the formulation)
has a higher order nonzero amplitude, since it must
satisfy a mass conservation relation which, to lowest order, reads:
\be
\delta_0 = - \frac{1}{2R} \sum_{k \neq 0} \mid \delta_k  \mid ^2.
\label{deze}
\ee

Following this observation, we split the deviation from $R$
in two terms, so that the radial function reads $r = R + \tilde{r} +
r_{0}$, where $r_0$ represents the zero mode (an expansion or
contraction of the circle), and $\tilde{r}$ the other modes.  We
scale them in the form $\tilde{r} \rightarrow w\tilde{r}$,
$r_{0} \rightarrow (w^2/R)r_{0}$, and write the position of the
interface $r(\phi)$ as:
\be
r(\phi) \rightarrow R( 1 + \varepsilon r(\phi) ),
\label{esca}
\ee
where:
\be
r(\phi)=\tilde{r}(\phi) +\varepsilon r_{0}.
\label{esco}
\ee
To simplify the notation, we have dropped out the time dependence.

The velocities will be scaled by:
\be
V_0 = \frac{1}{\mu_2 + \mu_1} \left(\frac{ Q (\mu_2 -\mu_1) }{2
\pi R} - \frac{b^2}{12} \Omega^2 (\rho_2 - \rho_1) R \right),
\ee
implying that time will be scaled by $R/V_0$.
Once equations (\ref{imp}), (\ref{impd}), (\ref{gmvip}), and
(\ref{nvir}) are made dimensionless, we have:
\be
\tilde{\gamma} = 2B\frac{\kappa_{\phi}}{1+\varepsilon r} + 2A\vec{U}
\cdot (\frac{\varepsilon r_{\phi}}{1+\varepsilon r} , 1 ) +
2\varepsilon \left(\frac{C}{(1+\varepsilon r)^{2}} - D \right)
r_{\phi},
\label{qp3}
\ee
for the vorticity, where:
\be
\begin{array}{c}
 \displaystyle{ B=\frac{b^2}{12} \frac{\sigma}{(\mu_1 + \mu_2) V_0 R^2}, \qquad
C=\frac{QA}{2 \pi R V_0}, \qquad  D=\frac{b^2}{12} \frac{ \Omega
^2 \triangle \rho R}{(\mu_1 + \mu_2) V_0},} \vspace{0.5cm} \\
\displaystyle{ \kappa=\frac{(1+\varepsilon r)^2 + 2 \varepsilon ^2
r_{\phi}^2 -\varepsilon (1+\varepsilon r)r_{\phi \phi}}{\left[
(1+\varepsilon r)^2 + \varepsilon ^2 r_{\phi}^2\right]
^{\frac{3}{2}}} , \qquad
A=\frac{\mu_{2}-\mu_{1}}{\mu_{2}+\mu_{1}}, \qquad \triangle \rho =
\rho_2- \rho_1.} \label{kappa}
\end{array}
\ee
For the integrals:
\be
U_{\hat{r}}(\phi) = \frac{1}{4 \pi} P \int_{0}^{2\pi} \frac{1 +
2\varepsilon r_2 + \varepsilon^2r_2^2}{\tan \left(\frac{\phi_2 -
\phi_1}{2}\right) \left[1 + \varepsilon f(\phi_1,\phi_2) +
\varepsilon^2 g(\phi_1,\phi_2)\right]}\tilde{\gamma} (\phi_2 , t)
d\phi_2,\label{qp1}
\ee
\be
U_{\hat{\phi}}(\phi) = \frac{1}{4 \pi} P \int_{0}^{2\pi} \frac{1 +
\varepsilon f(\phi_1,\phi_2) + \varepsilon ^2 r_1 r_2 - (1+
2\varepsilon r_2 + \varepsilon ^2r_2 ^2) \cos (\phi_2 - \phi_1) }{2
\sin ^2 \left(\frac{\phi_2 -\phi_1}{2}\right) \left[1 + \varepsilon
f(\phi_1,\phi_2) + \varepsilon^2
g(\phi_1,\phi_2)\right]}\tilde{\gamma} (\phi_2) d\phi_2, \label{qp2}
\ee
with:
\be
f(\phi_1,\phi_2) = r_1 + r_2, \qquad  g(\phi_1,\phi_2)= \frac{(r_1 -
r_2)^2}{4 \sin^2 \left(\frac{ \phi_2 -\phi_1 }{2}\right)} + r_1r_2,
\ee
and for the evolution of the deviation from $R(t)$:
\be
\varepsilon \frac {d\tilde{r}}{dt} + \varepsilon^2 \frac{dr_{0}}{dt}=
U_{\hat{r}} - \varepsilon \frac{r_{\phi} U_{\hat{\phi}}}{1 +
\varepsilon r}  + \frac{Q}{2 \pi R V_0} \left[ \frac{1}{1 +
\varepsilon r} - ( 1 - \varepsilon^2 r_{0})\right]. \label{popoito}
\ee

Eq.(\ref{popoito}) above is the only one which is sensitive to the
different possible scalings of the deviation from circularity,
due to the time derivatives.  The choice
we propose is the only one which is truly systematic.  Other
possibilities are not incorrect but will mix different orders of the
weakly nonlinear expansion at a given order in $\varepsilon$.

\subsection{The linear dispersion relation}

Our goal now is to obtain the linear dispersion relation and the
leading weakly nonlinear corrections, in both real and Fourier
space, and to discuss the interplay between rotation and injection
at the early stages of the instability.

We will use the following definitions for the average of a function
and for the Hilbert transform in the unit circle:
\be
\overline{f}  = \frac{1}{2\pi} \int_{0}^{2\pi} f(\phi) d\phi, \qquad
H[f(\phi^{\prime})] = \frac{1}{2\pi} P \int_{0}^{2\pi}
f(\phi^{\prime}) \cot \left(\frac{\phi^{\prime} - \phi}{2}\right)
d\phi^{\prime}.
\ee
The zero order of the vorticity and the two
components of the velocity are:
\be
\tilde{\gamma}^{(0)} = 2AU_{\hat{\phi}}^{(0)},  \qquad
U_{\hat{\phi}}^{(0)}=\frac{\overline{\tilde{\gamma}^{(0)}}}{2},
 \qquad U_{\hat{r}}^{(0)}= \frac{1}{2} H
[\tilde{\gamma}^{(0)}(\phi^{\prime})],
\ee
and thus:
\be
\tilde{\gamma}^{(0)}= A \overline{\tilde{\gamma}^{0}}.
\ee
This is different from its counterpart in the channel geometry.  Here
the equations for the vorticity at consecutive orders require the
knowledge of the average vorticity.  The
equations are solved by averaging on both sides.  For example, at zero order,
$A\neq1$ will lead to $\tilde{\gamma}^{(0)}=0$, but for $A=1$,
$\tilde{\gamma}^{(0)}$ can be any constant.  The presence of an
arbitrary constant is not a problem, since it is eliminated at
each order in the equation for the evolution of the interface.

Expanding the equations up to first order, and using
$\tilde{\gamma}^{(0)}=0$, the linearized equation for the evolution
of the interface deviation reads:
\be
\frac{d\tilde{r}}{dt}=U_{\hat{r}}^{(1)} - \frac{Q}{2\pi R
V_0}\tilde{r}= \frac{1}{2}H[\tilde{\gamma}^{(1)}] - \frac{Q}{2\pi R
V_0}\tilde{r},
\ee
with:
\be
\tilde{\gamma}^{(1)}= -2B(\tilde{r}+\tilde{r}_{\phi\phi})_{\phi} +
2(C-D)\tilde{r}_{\phi}.
\ee
We can now perform a Fourier transform of the equations and, after
reintroducing the adequate dimensions, we find:
\be
\lambda (n)= \left(\frac{b^2}{12}\frac{\Omega^2 \triangle \rho}{\mu_1
+ \mu_2} - \frac{Q A}{2 \pi R^2}\right)\mid n \mid - \frac{b^2}{12
R^3}\frac{\sigma}{\mu_1 + \mu_2}\mid n \mid (n^2-1) - \frac{Q}{2\pi
R^2} \label{lrr}.
\ee
This is the linear dispersion relation found in
\cite{Carrillo96}.

We now return to the question of the different
scaling alternatives introduced at the end of Section
\ref{dimeq-in-rad}.  Had we used the scaling $\varepsilon=w/R^2(t)$, the
term $Q/(2 \pi R^2)$ in (\ref{lrr}) would have disappeared before
reintroducing the dimensions.  On the contrary, for $\varepsilon=w$ this
term becomes multiplied by a factor $2$.  These changes in the
dimensionless linear dispersion relation come from the fact that a mode
$\delta_{n}$ in the scaling adopted in (\ref{esca}) and (\ref{esco})
becomes $\delta_{n}/R$ in the first of the scaling alternatives above.

The term $Q/(2 \pi R^2)$ at the end of Eq.  (\ref{lrr}) is
stabilizing for positive injection rate ($Q>0$) and destabilizing
for $Q<0$.  In a sense this is a purely geometric effect, which
contributes to the growth or decay of modes due to the expansion
or contraction of the base state.  As pointed out in
\cite{Rocco00} the term $Q/(2 \pi R^2)$, which can be scaled out,
must be distinguished from the rest, which do describe the
intrinsic instability of the problem.  Disregarding that last
term, it is then clear that injection and rotation ($CV_{0}$ and
$DV_{0}$ respectively) play an equivalent role in the linear
dispersion relation, since both appear multiplying $n$.  By
choosing the proper viscosity and density contrasts, they can
produce exactly the same stabilizing or destabilizing effect.
This is the counterpart, in the radial geometry, of the
equivalence between the roles of injection and gravity in the
channel geometry.  In the channel geometry, however, the
equivalence is exact and can thus be extended to all the nonlinear
evolution (in the appropriate reference frame).  This is not the
case in radial geometry, as shown in the next section.

\subsection{Weakly nonlinear theory with rotation}

The purpose of this section is to derive the leading order
nonlinear contributions for the general problem with both rotation
and injection.  We first recall that mass conservation related the
zero mode to the other modes.  In the original nonscaled variables
this reads:
\be
\dot{\delta}_0 = - \frac{Q}{2 \pi R^2}\delta_0 - \frac{1}{R}\sum_{k
\neq 0} \mid \delta_k \mid ^2 \lambda(k). \label{delzero}
\ee
To reproduce this result and the nonlinear couplings for the
other modes we start with the equation for the interface at order
$\varepsilon^2$:
\be
\frac{d\tilde{r}}{dt} + \varepsilon \frac{dr_{o}}{dt}= {\rm \cal{
O}}(\varepsilon^{0}) + \varepsilon \left(U_{\hat{r}}^{(2)} +
\frac{Q}{2\pi R V_0}\tilde{r}^2\right),
\label{newman}
\ee
where we have used that $U_{\hat{\phi}}^{(1)}= U_{\hat{\phi}}^{(0)} =
U_{\hat{r}}^{(0)} = \tilde{\gamma}^{(0)} =
\overline{\tilde{\gamma}^{(0)}} = 0$.  The velocity
$U_{\hat{r}}^{(2)}$ can be obtained from the expansion of
$U_{\hat{r}}$:
\be
U_{\hat{r}}^{(2)} = \frac{1}{2}H[
\tilde{\gamma}^{(2)}(\phi^{\prime})] +\frac{1}{2} H[(\tilde{r}(\phi
^{\prime}) - \tilde{r}(\phi))\tilde{\gamma}^{(1)} (\phi^{\prime})],
\label{vrdos}
\ee
where:
\be
\tilde{\gamma}^{(2)} = 2B\kappa ^{(2)} - 2B\tilde{r}\kappa^{(1)} +
2A \tilde{r}_{\phi}U_{\hat{r}}^{(1)} + 2 AU_{\hat{\phi}}^{(2)}
-4C\tilde{r}\tilde{r}_{\phi}.
\ee
$\kappa ^{(2)}$ and $U_{\hat{\phi}}^{(2)}$  can be computed from
(\ref{kappa}) and (\ref{qp2}) respectively.  The latter involves only
$\tilde{\gamma}^{(1)}$.  We obtain finally:
\be
\begin{array}{cc}\displaystyle{  \frac{d\tilde{r}}{dt} + \varepsilon
\frac{dr_{o}}{dt} = {\rm \cal{O}} (\varepsilon^0) +  \varepsilon } &
 \left( H \left[ B \kappa_{\phi^{\prime}}^{(2)} -
\frac{C+D}{2} (\tilde{r}^2)_{\phi^{\prime}} \right]  +
\tilde{r}H[s_{\phi^{\prime}}] + A H[H
[(\tilde{r}s_{\phi^{\prime\prime}})_{ \phi^{\prime\prime}}]] \right.
\\ & -  \left. A H \left[\tilde{r}(\phi^{\prime}) H[s_{\phi^{\prime\prime}
\phi^{\prime\prime}}] +
\tilde{r}_{\phi^{\prime}}H[s_{\phi^{\prime\prime}}] \right]
+\frac{Q}{2 \pi R V_0} \tilde{r}^2 \right),  \label{urdos}
\end{array}
\ee
with:
\be
\kappa^{(2)}= \tilde{r}^2 + \frac{\tilde{r}_{\phi}^2}{2}
+2\tilde{r}\tilde{r}_{\phi\phi}, \qquad s_{\phi}=\left( B(\tilde{r}
+\tilde{r}_{\phi\phi}) + (D-C)\tilde{r} \right)_{\phi}.
\ee

By introducing a superposition of Fourier modes in $\tilde{r}$ we
directly recover Eq.(\ref{delzero}).  For the other modes ($k \neq
0$) we get the result:
\be
\label{wnlrot}
\dot{\delta}_n = \lambda(n)\delta_n + \sum_{k \neq 0,n} \delta_{k}
\delta_{n-k}(F(k,n) + S(k,n) + \lambda(k) J(k,n)),
\ee
where:
\be
F(k,n) = \frac{\mid n \mid}{R} \left[ - \frac{Q A}{2 \pi R^2}
\left(\frac{1}{2} - \mbox{sgn}(kn)\right)  +
\frac{b^2}{24}\frac{\Omega^2 \triangle \rho}{\mu_1 + \mu_2}
\right],
\ee
\be
S(k,n)= \frac{\mid n \mid}{R} \left[ \frac{b^2}{12
R^3}\frac{\sigma}{\mu_1 + \mu_2}\left(1 - \frac{k}{2}(n +
3k)\right) \right],
\ee
\be
J(k,n) =-\frac{1}{R}\left[A\mid n \mid(1 - \mbox{
sgn}(kn))+1\right]. \ee For the particular case $\Omega = 0$, the
expression above reproduces the result of Miranda and Widom
\cite{Mirandar98}.  We find that the presence of rotation does not
change the formal structure of the equations, but introduces new
terms.

To emphasize the roles of injection and rotation, we define the
quantity $H(k,n)$ as the part of the coupling matrix in the r.h.s.
of Eq.(\ref{wnlrot}) which contains $Q$ and/or $\Omega$.  This
quantity reads:
\be
H(k,n)=\frac{\mid n \mid}{2R} \left( \frac{Q A}{2 \pi R^2}+
\frac{b^2}{12}\frac{\Omega^2 \triangle \rho}{\mu_1 + \mu_2}\right) +
J(k,n)\mid k \mid \left(\frac{b^2}{12} \frac{\Omega^2\triangle
\rho}{\mu_1 + \mu_2} - \frac{QA}{2 \pi R^2}\right) + \frac{Q}{2 \pi
R^3}.
\ee
We observe that experimental parameters occur only in two groups,
namely:
\be
\tilde{\Omega}=\frac{b^2}{12} \frac{\Omega^2\triangle \rho}{\mu_1
+ \mu_2}, \qquad \tilde{Q}=\frac{QA}{2 \pi R^2}. \ee Both of them
turn out to multiply the same functions of the wave numbers,
except for changes in sign.  However, notice that the relative
sign of $\tilde{\Omega}$ and $\tilde{Q}$ is different in the
linear part and in the leading nonlinear part of the dynamics.
This clearly shows that the effects of injection and rotation
cannot be interchanged by simple changes in parameters, and that
the intrinsic new dimensionless parameter related to rotation
already shows up at the early nonlinear regime.

It is also remarkable that, for $Q \neq 0$, $R(t)$ is time dependent,
and the relative weight of $\tilde{\Omega}$ and $\tilde{Q}$ changes with
time.  This may have remarkable consequences in the highly nonlinear
regime.  For instance, rotation can prevent the formation of finite time
singularities in the case with zero surface tension \cite{Rocco00}.

Finally, notice that at nonlinear order there is also a
geometrical term, $Q/(2 \pi R^3)$, independent of $n$.  This
term cannot be scaled out by the same procedure that eliminated its
counterpart at linear order.

\section{Numerical study of an exact solution and its weakly nonlinear
approximation} \label{Sec:5}

The purpose of this section is to put the weakly nonlinear analysis to
the test, as a quantitative approximation.  We will check it against an
exact time dependent solution of the case $B=0$ which evolves towards
the Saffman--Taylor finger of width $\lambda=1/2$ in the channel
geometry.  This will allow us to check how fast is the convergence of
the expansion and how accurate it may be even beyond its radius of
convergence, when it becomes an asymptotic series.

We define $F$ as the ratio between the maximum height of the interface
(at the finger tip) and half the width of the cell (in our case
$L=2\pi$).  $F$ is a dimensionless amplitude which is proportional to
$\varepsilon$ and thus measures how deep the system is into the
nonlinear regime.  Furhermore, since the system is described by the
evolution of a curve, it may also be convenient to have a more global
characterization of its configuration in order to compare the exact
result and the different approximations.  We propose the use of the flux
$\Phi$, defined as the total amount of fluid 1 per unit length and unit
time flowing across the horizontal line located at the mean interface
position \cite{Magdaleno00}).

\subsection{The time dependent exact solution with $B=0$
and $\lambda=1/2$}

An explicit solution of the problem without surface tension
($B=0$) and valid for any viscosity contrast $A$, describing the
growth of a single finger which asymptotically fills half
of the channel ($\lambda=1/2$), can be obtained from
Eqs.(\ref{mappi}), (\ref{setun}), and (\ref{setdos}).  This reads:
\be
f(w,t)= -\ln w +  V_{o}t + a(0) + \ln \left[\sqrt{1+
(b(0)^2-1)\exp(2V_{o}t)} +w\right],
\label{fwt}
\ee
with $a(0)$ and $b(0)$ as initial constants, and $b(0)\gg 1$.  Since
this solution is valid for any $A$, particularly $A=0$, all terms of
the weakly nonlinear equation (\ref{generity}) depending on $A$
will necessarily have no contribution.

\begin{figure}
\begin{center}
\epsfig{file=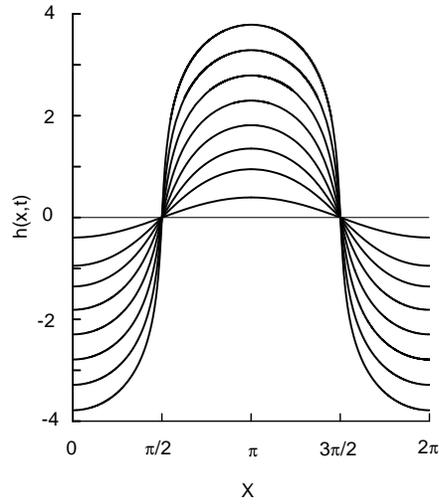,width=6cm} \vspace{0.1cm}
\caption{Evolution of the interface at different values of $\pi
F$.} \label{FigEx1}
\end{center}
\end{figure}

\begin{figure}

\begin{center}
\epsfig{file=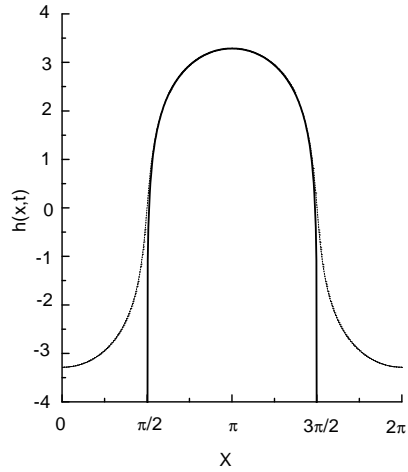,width=6cm} \vspace{0.5cm}
\caption{Comparison of the exact solution
(Eq.(\protect{\ref{fwt}})) at $F=1$ (thin line) and the
Saffman-Taylor stationary finger (thick line).} \label{FigEx2}
\end{center}
\end{figure}

We will take the following values for the parameters:
\be
V_{o}=\frac{1}{2}, \qquad b(0)=\frac{1}{\nu}=1000, \qquad a(0)= \ln
\nu +\frac{\nu^2}{2}.
\ee
Fig.\ \ref{FigEx1} shows the evolution of the interface for this
solution.  When $F\simeq1$ the finger shape is indistinguishable
from the stationary Saffman--Taylor solution
(Fig.\ \ref{FigEx2}),
except near the points $x=\pi/2$ and $x=3\pi/2$
when the asymptotes of the finger must develop at infinite
time.  When $F = 1/\pi$, the slope of the interface
becomes $1$ at these two points and the series (\ref{generity})
loses its convergence, according to the discussion of Section
\ref{SectionIIIA}.

\begin{figure}
\begin{center}
\epsfig{file=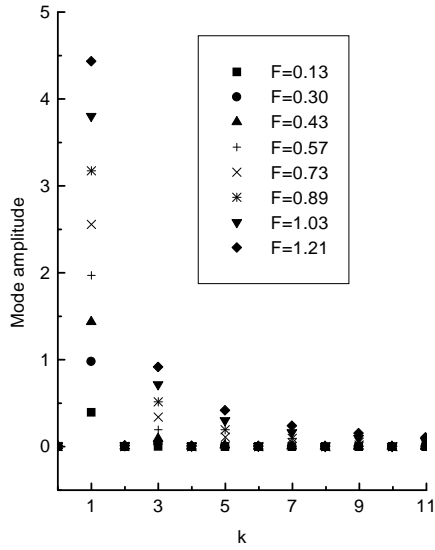,width=6cm} \vspace{0.5cm}
\caption{Amplitude of the Fourier modes of the exact solution at
consecutive values of $F$.} \label{FigEx3}
\end{center}
\end{figure}

In Fig.\ \ref{FigEx3} we present the amplitude of the Fourier
modes of the exact solution.  It is clear that relatively few
modes suffice to accurately reproduce the stationary finger shape
in the tip region.

\subsection{Expansion of the exact solution}

\begin{figure}
\begin{center}
\epsfig{file=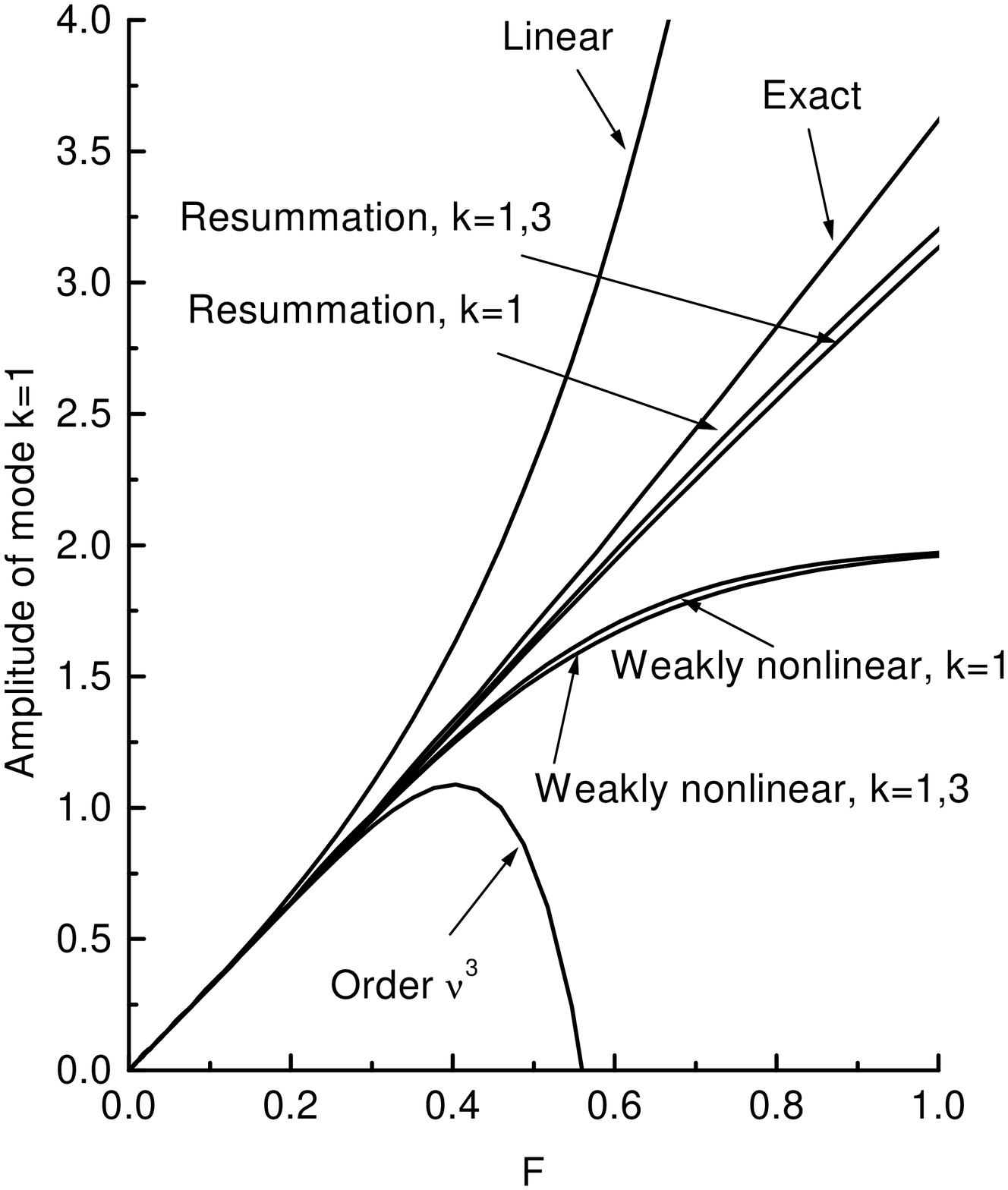,width=6cm} \vspace{0.5cm} \caption{The
first mode amplitude as a function of the factor F for different
approximations (see text for details).} \label{FigFp3}
\end{center}
\end{figure}

\begin{figure}
\begin{center}
\epsfig{file=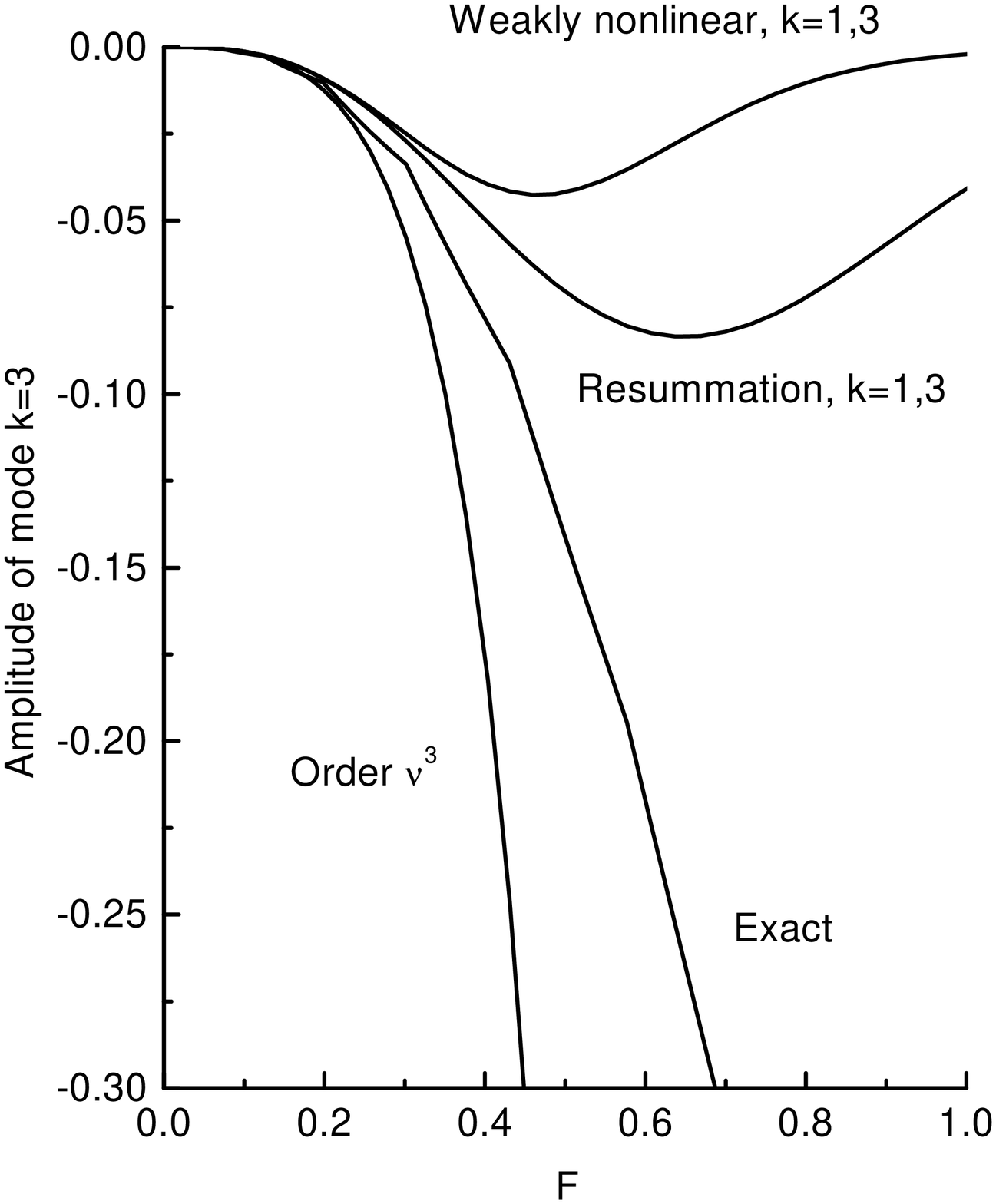,width=6cm} \vspace{0.5cm} \caption{The
third mode amplitude as a function of the factor F for different
approximations (see text for details).} \label{FigFp4}
\end{center}
\end{figure}

In Section \ref{subsecdit} we expanded the exact solution
(\ref{mappi}) in a small parameter $\nu$.  This expansion revealed
a hierarchy of amplitudes of the different modes
for an initial condition close
to the planar interface.  The
solution (\ref{fwt}) can be expanded similarly and yields:
\be
\nu = \frac{1}{b(0)}, \ \ \ \alpha (t) = - \frac{\nu}{b(t)}, \ \ \
  d(t)=V_{o}t + \ln b(t), \ \ \ b(t)= \sqrt{ \nu^2 + (1 -
\nu^2) \exp(2V_{o}t)}.
\ee
In agreement with (\ref{hexpan}), and using
the same parameters as in the previous section, we obtain up to third
order:
\be
h(x,t)=\nu e^{t/2} \cos x  + \nu ^3
\left[\left(\frac{1}{2} e^{t/2} - \frac{1}{8}
e^{3t/2}\right)\cos x - \frac{1}{24} e^{3t/2} \cos 3x \right].
\ee
The exact flux $\Phi$ can be easily computed from
the stream function, which in its turn can be related to the form of
the mapping Eq.(\ref{fwt}) (see for instance
Ref.\cite{Magdaleno00}).  In Fig.\ \ref{FigFp1} we compare the
flux $\Phi$ of the approximate result above with the exact result
coming from (\ref{fwt}).  We clearly observe that the first
correction to the linear regime improves the flux (to a 5 \%
accuracy) from $ F \simeq 0.12 $ to $ F \simeq 0.17$.  The
amplitude of the mode $k=1$ (Fig.\ \ref{FigFp3}) is fairly
well reproduced
by the linear approximation up to $F \simeq 0.20$, and the correction
of order $\nu^3$ gives a value for the mode $k=3$ reasonably good
up to $F \simeq 0.20$ (Fig.\ \ref{FigFp4}) and improves the mode $k=1$
almost until $F \simeq 0.25 $.  Above these values of $F$ the
deviation from the exact interface is exponential.

Notice that the expansion up to $\nu^3$ considered here does not
account for the complete three mode coupling of
Eq.(\ref{triorder}) which contains part of the higher orders in
$\nu$.  This explains why going from order $\nu$ (linear) to order
$\nu^3$ improves only slightly the shape of the finger, the flux $\Phi$,
and the amplitude of the mode $k=1$.

\subsection{Weakly nonlinear expansion}

In this section we study the mode coupling equation
(\ref{triorder}).  We solve this equation in cases where only
modes $k=1$ and $k=1,3$ are present, and compare the result with
both the exact solution and the approximation to order $\nu^3$
introduced in the two previous sections.

\begin{figure}
\begin{center}
\epsfig{file=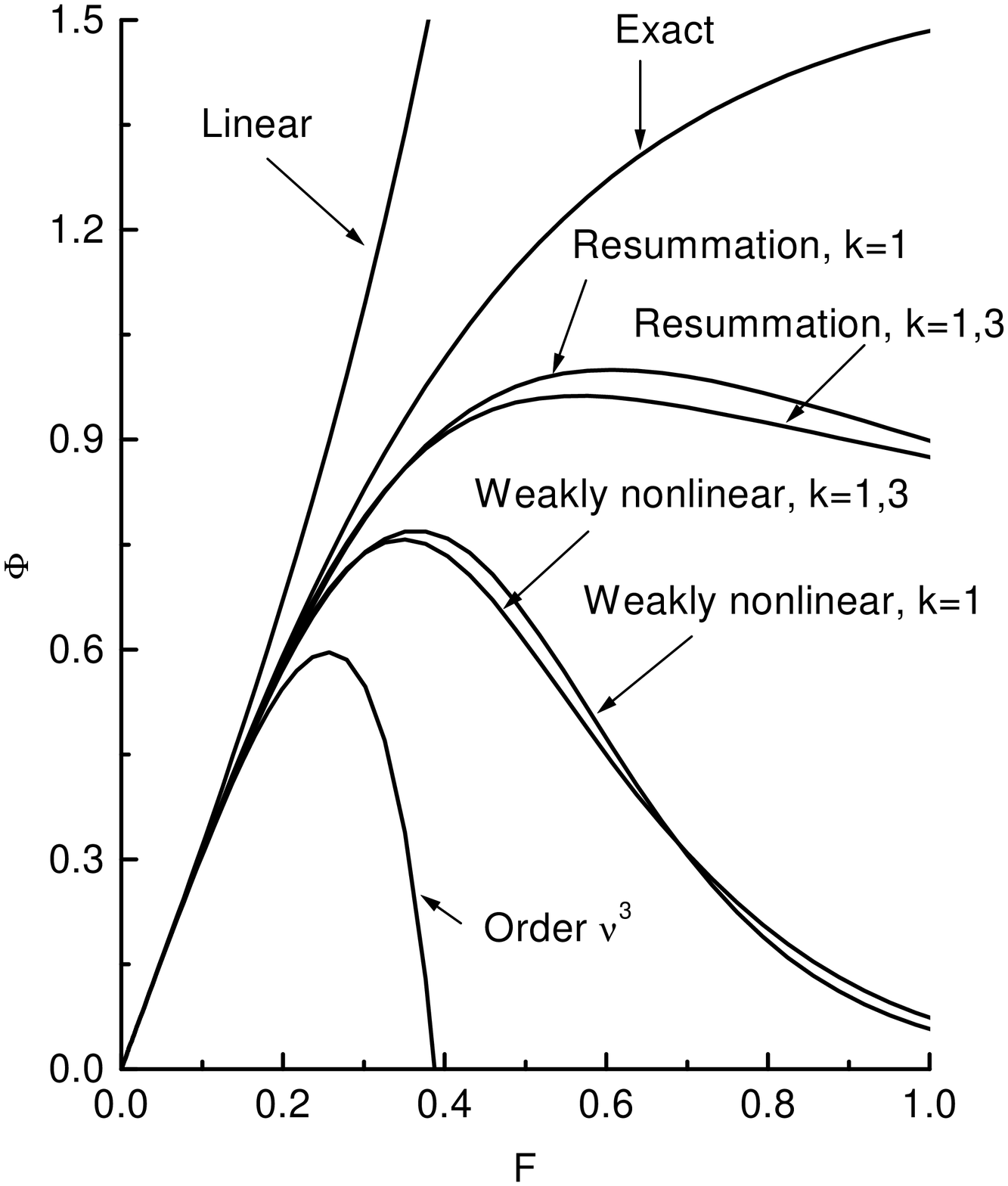,width=6cm} \vspace{0.5cm} \caption{Flux
$\Phi$ as a function of the factor F for different approximations
(see text for details).} \label{FigFp1}
\end{center}
\end{figure}

\begin{figure}
\begin{center}
\epsfig{file=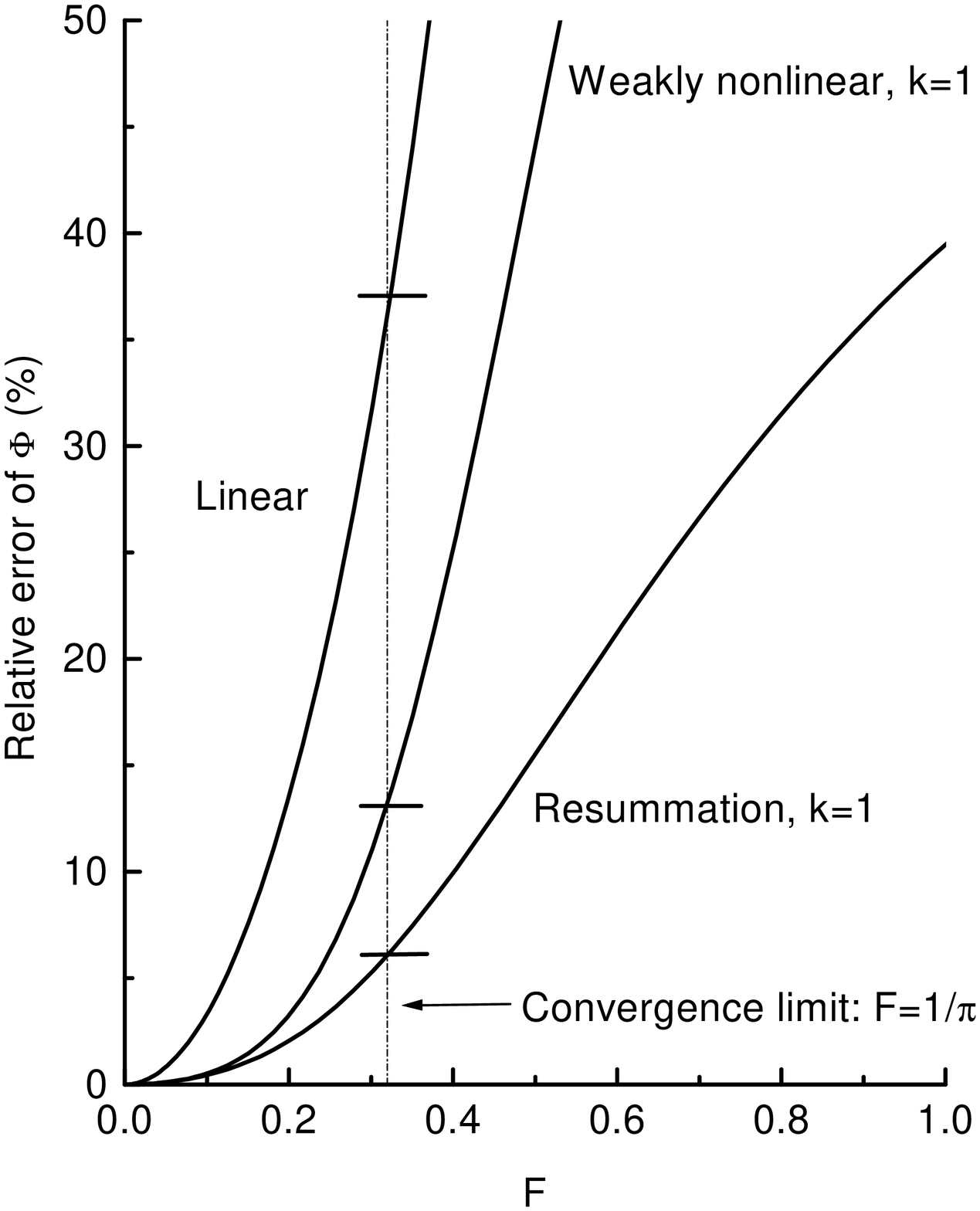,width=6cm} \vspace{0.5cm}
\caption{Relative error of the flux $\Phi$ as a function of the
factor F for different approximations (see text for details).}
\label{FigFp2}
\end{center}
\end{figure}

We consider the initial condition $\beta
_{1}(0)=\varepsilon=0.001$ for the first mode and zero for the
rest of modes.  We use Eq.\ (\ref{triorder}) recalling that, since
$\beta _{i}$ represent the amplitudes of the cosine functions,
$\delta_{i} = \beta_{i}/2$.  The result is that the modes
$k\neq 1$ remain zero and:
\be
\beta _{1} (t)=\frac{\varepsilon \exp(0.5t)}{\sqrt{1+ 0.25
\varepsilon ^2 (\exp(t) -1)}}, \ee where we see that the mode
$k=1$ saturates to a finite value as $t \to \infty$.  Returning to
Fig.\ \ref{FigFp1} and \ref{FigFp2}, we see that the flux $\Phi$
approximates the exact value to a 5 \% uncertainty) until $F
\simeq 0.23$.  Fig.\ \ref{FigFp3} shows that the amplitude of the
first mode starts deviating significantly from the exact solution
around $F \simeq 0.40$.

Next, we add the third mode to the equation (\ref{triorder}) with
the initial condition $ \beta_{1}= \varepsilon + 3 \varepsilon
^3/8$ and $\beta_{3}= \varepsilon ^3/24$.  The set of equations
to be solved numerically is:
\be
&& \dot{\beta}_{1}=\frac{1}{2}\beta_{1} - \frac{1}{8}\beta_{1}^3 -
\frac{5}{4} \beta_{3}^2 \beta_{1} + \frac{1}{4} \beta_{1}^2
\beta_{3},
\\ && \dot{\beta}_{3}=\frac{3}{2} \beta_{3} - \frac{3}{4} \beta_{1}^2
\beta_{3} - \frac{27}{8} \beta_{3}^3. \label{system} \ee As shown
in Figs.\ \ref{FigFp1} and \ref{FigFp4} the flux does not improve
significantly, while the amplitude of the mode $k=3$ remains
acceptable almost until $F\simeq 0.25$, and improves the previous
result by more than a 50 \% almost until $F\simeq 0.33$.  Beyond
this value of $F$ the third mode falls to zero due to the second
term in the r.h.s. of Eq.(\ref{system}), proportional to
$\beta_{1}^2 \beta_{3}$ but negative.  This is a signature that
higher order terms, for instance of order $\beta_{1}^4 \beta_{3}$,
become important when $\beta_{1}$ is of order one.

From this study we can conclude that using only two modes and the lowest
order nonlinear correction for this case (involving three mode
couplings), both the shape of the interface and the total flux are
reasonably well described within the whole range of convergence of the
series (up to $F \simeq 1/\pi$).  This indicates that the weakly
nonlinear description of the problem works remarkably well at the
quantitative level, at least in some physically relevant cases.  To what
extent this conclusion holds for more complicated situations, for
instance those involving competition of different fingers, remains an
open question.

\subsection{Partial resummation}

According to Eq. (\ref{triorder}), we can write a closed equation for
the mode $k=1$ which is correct up to third order couplings, of the
form:
\be
\dot{\beta}_{1} = \frac{1}{2} \beta_{1} - \frac{1}{4} \beta_{1}^2
\dot{\beta}_{1},
\label{laderiva}
\ee
where the mode $k=3$ has been set initially to zero.

The presence of $\dot{\beta}_{1}$ in the right hand side gives
rise to terms of order higher than $\beta_{1}^3$.  Thus, solving
this equation amounts to doing a partial resummation of these
higher order terms.  This kind of closed differential equation was
already obtained in Ref.\cite{Mirandac98}.  In our
approach this differential equation is obtained in a systematic way by
identifying the terms $\lambda(n) \beta_{n}$ and replacing them
for $\dot{\beta}_{n}$.  For example, in Eq. (\ref{triorder}) the
term $T(k,s,l)$ could have been partially resummated in the
two mode coupling differential equation.

The solution of Eq.(\ref{laderiva}) yields the transcendental
form:
\be
t = 2 \ln \left(\frac{\beta_{1}}{\varepsilon}\right)
+\frac{1}{4}(\beta_{1}^2-\varepsilon^2).
\ee
A numerical computation of the flux based on this equation shows a
dramatic improvement up to $F \simeq 1/\pi$.  It is
surprising and remarkable that the equation for a single mode,
including the resummation suggested by the first nonlinear
correction, reproduces the exact solution to a great accuracy in
the full range of convergence and defines an excellent
approximation further inside the nonlinear regime.  This suggests
that this kind of resummation is not arbitrary, and might have a deeper
physical justification.

In addition, we obtain an amplitude for the mode $k=1$ within 10 \%
accuracy up to $F \simeq 0.8$, and it is still quite good until
$F=1$.  Although the exact spectrum shows the increasing
importance of the mode $k=3$ at these values of $F$, the agreement
between the exact result and this approximation for a single mode
$k=1$ is quite remarkable.

We recall that the series converges up to $F \simeq
1/ \pi$. Beyond this value of $F$, increasing the order of
mode coupling in Eq. (\ref{laderiva}) does not guarantee that the
approximation improves.  Accordingly, there is an optimal finite
order of approximation for a given value of $\varepsilon$, as
usual in asymptotic series.  In this sense, our results above seem
to indicate that Eq.(\ref{laderiva}) is the optimal approximation
for the mode $k=1$ in the asymptotic (nonconvergent) region (i.e.,
for moderately long times).

In the same spirit of Eq. (\ref{laderiva}), we can  also derive a
closed system of differential equations which includes third order
couplings for the modes $k=1,3$.  Its solution shows no
qualitative differences with the case of the single mode $k=1$
above, as far as the flux and the amplitude of the first mode are
concerned.  The improvement in the amplitude of the third mode and
consequently in the shape of the finger does not reach the high
values of $F$ that the first mode reaches ($F \simeq 0.8$), as we
can see in Fig.\ \ref{FigFp4}.  In a short range of $F$ the mode
$k=3$ is better than the case shown in the previous section, but
deviates from the exact solution much earlier than the mode $k=1$
for the same reasons than in the previous section.  Specifically
the mode $k=3$ is extremely good until $F \simeq 0.25$, starts
deviating by more than a 10 \% at $F \simeq 0.33$ and is
about 50 \% of the value at $F \simeq 0.40$.

We conclude that the weakly nonlinear formalism, apart from its nominal
range of validity for very small amplitudes where it converges very fast
to the exact dynamics, can describe regimes beyond those
expected in principle with reasonable accuracy.  In the case of the
growth of a single finger,
for instance, we have explicitly seen that simply two modes together
with an appropriate partial resummation, yields a remarkably good
approximation in the full range of convergence of the expansion ($F <
1/\pi$).

\section{Conclusion and perspectives}
\label{conclusions}

We have developed a systematic scheme to derive the successive orders of
mode couplings in a weakly nonlinear regime, adapted to the study of
interfacial dynamics in Hele--Shaw flows.  This has been done with full
generality, including both the channel geometry driven by gravity and
pressure, and the radial geometry with arbitrary injection and
centrifugal driving, which includes the case of a rotating Hele--Shaw
cell.  The method could also be applied to even more general
(position dependent) drivings.  The formulation in real space has
enabled us to address the issue of convergence of the mode coupling
expansion.  We have found that the exact condition for convergence in
the channel geometry is $| h_x | < 1$ in every point of the interface.

We have carried out the explicit derivation of nonlinear couplings up to
third order in the case of channel geometry, in order to obtain the
leading nonlinear contributions in cases where the second order
contribution vanishes.  These include the case of zero viscosity
contrast and the time dependent single finger solution of width
$\lambda=1/2$.

On the analytical side, we have also illustrated the usefulness of
the weakly nonlinear analysis in elucidating the role of the
different parameters on the dynamics, in some examples.  In the
case of single finger solutions growing in a channel, we have
shown how an exact nontrivial property of the problem (the
relationship between $\lambda$ and viscosity contrast $A$ for zero
surface tension) can be extracted from an analysis order by order,
without really knowing the exact solution.  In the case of
rotating Hele--Shaw flows, we have discussed the asymmetry between
the roles of injection and rotation at the lowest nonlinear order,
as opposed to what happens in the channel geometry, where the
equivalence between gravity and injection driving is exact.

On the numerical side we have checked the predictions of the
weakly nonlinear analysis, at different orders, against exact
solutions.  We have found that the convergence is quite fast for
small amplitudes.  Furthermore, in the case of a single finger
growing in a channel we have explicitly seen that the analysis to
low orders (up to three mode couplings) yields a good description
of the interface dynamics even close to the radius of convergence.
With an appropriate resummation scheme, the prediction is
relatively good even beyond that point.

Some of the most interesting applications of the systematic weakly
nonlinear approach have not been developed here since they deserve
a separate, in--depth analysis.  One of them is the study of the
scaling properties of fluctuations in stably stratified Hele--Shaw
flows with external noise \cite{porous}.  A more direct
application of our formalism which also deserves a separate
analysis is the derivation of amplitude equations for the region
near the instability threshold through a center manifold
reduction.  This would made contact with what is most commonly
referred to as ``weakly nonlinear'' approach in the literature of
pattern formation.  A detailed study of this point with a careful
discussion of the nature of the bifurcation and its implications
will be presented elsewhere \cite{Alvarez02}.

\section{Acknowledgements}
We acknowledge financial support from the Direcci\'on General de
Ense\~{n}anza Superior (Spain) under Projects PB96-1001-C02-02 and
PB97-0906, E. Alvarez--Lacalle also acknowledges a grant from the
Comissionat per a Universitats i Recerca of the Generalitat de
Catalunya.  The work was also supported by the European Commission
Project ERB FMRX-CT96-0085 (Training and Mobility of Researchers).

\newpage



\newpage

\end{document}